\documentclass[
nofootinbib,
superscriptaddress,amsfonts,amssymb,amsmath, twocolumn
]{revtex4-2}
\usepackage{amsmath,bm,mathtools,physics}
\usepackage{hyperref}
\usepackage{float}
\usepackage[utf8]{inputenc}
\numberwithin{equation}{section}
\renewcommand\theequation{\arabic{section}.\arabic{equation}}

\usepackage{perpage} 
\usepackage{mathrsfs}
\usepackage{subfigure}
\usepackage{array}
\MakePerPage{footnote}
\usepackage{tabularx}
\usepackage{hyperref}

\usepackage{graphicx}
\usepackage{comment}
\graphicspath{ {/} }

\newcommand{\V}{{\mathbf{{V}}}}

\newcommand{\Di}[1]{\prod_{i=1}^{#1} \frac{C_i}{\Gamma(\Delta_i)} \widehat{D}_{\Delta_i}^{M_iA_i}}

\begin{document}
\title{Mellin Amplitude for $n$-Gluon Scattering in Anti-de Sitter}
\author{Jinwei Chu}
\email[]{jinweichu@uchicago.edu}
\affiliation{Department of Physics, University of Chicago, Chicago, IL 60637, USA}
\author{Savan Kharel} 
\email[]{skharel@uchicago.edu}
\affiliation{Department of Physics, University of Chicago, Chicago, IL 60637, USA}

\begin{abstract}
In AdS/CFT, we introduce a robust method for computing $n$-point gluon Mellin amplitudes, applicable in various spacetime dimensions. Using the Mellin transform and a recursive algorithm, we efficiently calculate tree-level gluon amplitudes. Our approach simplifies the representation of higher-point amplitudes, eliminating the need for complicated integrations. Crucially, the resulting amplitudes closely mirror those in flat space, allowing a straightforward dictionary between the two settings circumventing explicit calculations.
\end{abstract}

\date{\today}
\maketitle

\section{Introduction}
In the context of Anti-de Sitter (AdS) spacetime, the natural generalizations of the scattering amplitudes are the correlation functions in the associated dual conformal field theory (CFT)~\cite{Maldacena:1997re,Witten:1998qj}. Some important strides have been made, revealing simplifications and elegant structures akin to those observed in flat space. A particularly useful approach in calculating AdS amplitudes is the adoption of the Mellin basis~\cite{Mack:2009mi,Penedones:2010ue}. This technique offers advantages analogous to those provided by the momentum basis in Minkowski space computations. In an important development, researchers have created a `dictionary' for scalar particles that translates the familiar Feynman rules from flat-space amplitudes to write Mellin amplitudes directly~\cite{Paulos:2011ie,Fitzpatrick:2011ia}. Complementing this development, there has been considerable progress in  AdS momentum space, as reflected in recent studies \cite{Bzowski:2015pba,Albayrak:2018tam,Isono:2018rrb,Farrow:2018yni,Albayrak:2019yve,Bzowski:2019kwd,Isono:2019ihz,Albayrak:2020isk,Bzowski:2020kfw,Jain:2020puw,Marotta:2022jrp}. However, despite these advancements, the field still faces a significant hurdle. The prime difficulty is the systematic computation of amplitudes that include external particles with spin, particularly as we attempt to move beyond the simpler cases, like three and four-point functions. (see some progress for spinning AdS amplitudes~\cite{Kharel:2013mka,Costa:2014kfa,Sleight:2017fpc,Nishida:2018opl,Albayrak:2019asr,Goncalves:2019znr,Alday:2022lkk,Bissi:2022mrs,Li:2023azu,Alday:2023kfm}.)

There are several compelling physical motivations for considering higher-point gluon AdS amplitudes: Firstly, higher-point gluon amplitudes in flat space, while appearing complicated, are often simpler as demonstrated by the Parke-Taylor formula~\cite{Parke:1986gb} and innumerable subsequent research (see for instance \cite{Witten:2003nn,Arkani-Hamed:2008owk}). Secondly, study of gluon scattering has catalyzed the development of on-shell recursion relations, enabling systematic amplitude computations based on fundamental three-point gluon amplitudes~\cite{Britto:2005fq}. Finally, these higher-point amplitudes provide essential building blocks for constructing more involved gravity amplitudes through techniques such as color-kinematic duality and double-copy construction~\cite{Bern:2010ue}.

This letter aims to construct the direct dictionary between the external $n$-gluon AdS amplitude and the flat-space amplitude. Some hints for the connection in lower-point gluons was observed in~\cite{Paulos:2011ie}. However, the complexities inherent in higher-point spinning amplitudes have stalled progress. In general bulk integration and propagation of indices make tree-level external gluon amplitudes for higher-point functions exceedingly difficult to compute. Hence, to date not much progress has occurred to investigate if this relationship (or the modification) can extend to $n$-point. In this letter, with the aid of novel and explicit computations we are able to see the map between flat space and AdS computations.

This letter is structured as follows: Section \ref{II} serves a pedagogical purpose, introducing the fundamental techniques employed in our analysis. We outline the concept of Mellin amplitudes, the factorization of bulk-to-bulk propagators, and the differential operator associated to the spinning boundary-to-bulk propagator. Additionally, we explore a concept analogous to momentum conservation for gluon AdS amplitudes, which streamlines our calculations. By combining these elements, we can formulate a general expression for a recursion relation. This provides a systematic approach to constructing an $n$-point amplitude from its lower-point counterparts and a three-point amplitude.

In Section \ref{III}, we introduce the concept of the {reduced Mellin amplitude} and proceed to simplify the previously mentioned recursion relation within the framework of pure Yang-Mills theory. We observe a significant simplification in our expression, which results from the antisymmetry between the two external legs in the three-point amplitude. This simplification enables us to more efficiently compute across various Feynman-Witten diagram topologies for gluons in AdS.

To validate the efficacy of our methodology, we have conducted calculations extending up to eight-point amplitudes in our accompanying paper, emphasizing the scalability and applicability of our approach~\cite{accompanying}. Surprisingly, we discovered a precise relation between the Mellin amplitudes and their flat-space equivalents for the cases we examined, which we elaborate on in Section \ref{IV}. Based on these promising findings, we hypothesize that our results could be extended to all tree-level gluon amplitudes. We present our final remarks in Section \ref{V}.

\section{Mellin amplitude and Recursion}
\label{II}
In our treatment, \text{AdS}$_{d+1}$ is envisioned as a hyperboloid embedded in a Minkowski space of $(d+2)$ dimensions, $\mathbb{M}^{d+2}$, with $X$ as point in the bulk AdS and $P$ as its boundary ($P^2=0$). Mellin space is the natural space for correlation functions in CFTs, especially for weakly coupled AdS duals~\cite{Mack:2009mi,Penedones:2010ue}. We can write down a conformal field theory spinning correlator in the Mellin representation as
\begin{equation}
\label{Mellin}
\langle J^{M_1} \cdots J^{M_n} \rangle=\int [d\gamma] \mathscr{M}^{\textbf{$\bm{M_i}$}}_n  \Gamma(\gamma_{ij})P_{ij}^{-\gamma_{ij}}\ ,
\end{equation}
where we have defined $ [d\gamma] \equiv \prod_{i<j}^n d\gamma_{ij}/2\pi i$. The Mellin amplitude, \mbox{$\mathscr{M}^{\textbf{$\bm{M_i}$}}_n \equiv \mathscr{M}_n^{M_1 \cdots M_n}$}, can depend on Mellin variables $\gamma_{ij}$ and boundary points $P_i$. Here, $P_{ij}=-2 P_i\cdot P_j$ and the scaling dimension for the spin-1 field is $\Delta_i=d-1$.

In AdS amplitude analysis, Witten diagram mirrors Feynman's, comprising  of vertices and propagators. Intriguingly, vector field's boundary-to-bulk propagators emerge from acting a differential operator on their scalar counterparts: \mbox{$\mathcal{E}_{\Delta_i}^{M_iA_i}(P_i,X)=\widehat{D}_{\Delta_i}^{M_iA_i}\mathcal{E}_{\Delta_i}(P_i,X)$}, with the scalar propagator \mbox{$\mathcal{E}_{\Delta_i}(P_i,X)=C_i (-2 P_i\cdot X)^{-\Delta_i}$} \footnote{Here, $C_i={\Gamma (\Delta_i)}/{(2\pi^h\Gamma \left(\Delta_i+1-h\right))}$ with $h=d/2$.}. The differential operator is given by~\cite{Paulos:2011ie}
\begin{equation}
\label{DMA}
 \widehat{D}_{\Delta_i}^{M_iA_i}=\frac{\Delta_i-1}{\Delta_i}\eta^{M_iA_i}+\frac{1}{\Delta_i}\frac{\partial}{\partial P_i^{M_i}}P_i^{A_i}\ .
 \end{equation}
 Crucially, the utilization of this operator greatly streamlines functional manipulations. Let us consider a function $F(P_i)$ with a scaling dimension  $\Delta_i-1$ in  $P_i$. One can see that,
\begin{equation}
\label{eq:eigD}
\widehat{D}_{\Delta_i}^{M_iA_i} \frac{\partial}{\partial P_i^{A_i}}F(P_i) =0\ .
\end{equation}
Here we present a simple illustrative example:
\begin{equation}
    \label{eigD1}
\widehat{D}^{M_iA_i} \sum_{j\neq i}P_{j,A_i}\prod_{k<l}\Gamma(\gamma_{kl})P_{kl}^{-\gamma_{kl}}=0\ .
\end{equation}
In the equation, we have shifted the Mellin variables so that each term has the same scaling dimension as $\Delta_k$, for any $P_k$. We will call this feature (and analog examples such as \ref{eigD2} etc.) \emph{generalized momentum conservation} as they help replace a boundary point by other boundary points.\footnote{Another example of applying (\ref{eq:eigD}) is that
\begin{equation}
\label{eigD2}
\widehat{D}_{\Delta_i}^{M_iA_i}\left(\eta_{A_iA_j}-2P_{i,A_j}\sum_{k\neq i}P_{k,A_i}\right)\prod_{l<m}\Gamma(\gamma_{lm})P_{lm}^{-\gamma_{lm}}=0\ .
\end{equation}
Here again each term has the same scaling dimension as $\Delta_k$ for any $P_k$.} We remind the reader that one uses such techniques, i.e. $\sum_j k_j = 0$ in flat-space scattering amplitude computations.

Bulk-to-bulk propagators represent exchange fields of scaling dimension $\Delta$ in the bulk. In the embedding formalism, the bulk-to-bulk propagator can be factorized into two boundary-to-bulk propagators integrated over an internal boundary point $Q$. For vector fields~\cite{Balitsky:2011tw}, 
\begin{equation}
\mathcal{G}^{AB}_\Delta(X_1,X_2)= \! \int_{c,\partial} \! \mathcal{E}^{MA}_{h-c}(Q,X_1)\eta_{MN} \mathcal{E}^{NB}_{h+c}(Q,X_2) \,
\end{equation}
where \mbox{$\int_{c,\partial} \equiv \int_{-i\infty}^{i\infty}\frac{dc}{2\pi i} \left(\frac{4c^2(h^2-c^2)}{\left(c^2-(\Delta-h)^2\right)^2}\right) \int_{\partial\text{AdS}} dQ$}. Here the expression has double poles in $c$, which makes the integral over $c$ complicated. Later, we will see that this pole simplifies in our computation. 

\begin{figure}
	\centering
\includegraphics[scale=0.7]{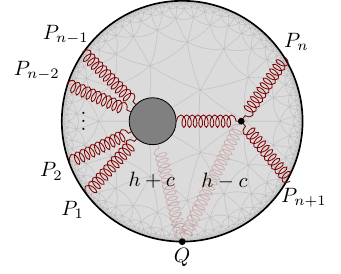}
\caption{\label{nchannel} Schematic of how $n+1$ Mellin amplitude factorizes into two sub-amplitudes.
}
\end{figure}

In our analysis, we will factorize an $(n+1)$-point amplitude into an $n$-point and a 3-point amplitude, as illustrated in Fig.~\ref{nchannel}.
In order to integrate over the internal boundary point $Q$, we take the following two important steps. First, we use the \emph{generalized momentum conservation} to substitute $Q$ with a free index with the other $P_i$'s. 

Next, we can utilize Symanzik's formula\footnote{Namely for $\sum_{i=1}^nl_i=d$,
\begin{multline}
\label{symanzik}
\int_{\partial\text{AdS}} dQ\ \Gamma(l_i)\left(-2P_i\cdot Q\right)^{-l_i}
=\pi^h\int [d\gamma_{ij}]\Gamma(\gamma_{ij})P_{ij}^{-\gamma_{ij}}\ .
\end{multline}}
to perform the integration over $Q$. After shift of $\gamma_{ij}\to\gamma_{ij}-\gamma_{ij}'$ for $1\le i,j\le n-1$ and $i,j=n,n+1$ (where $\gamma_{ij}'$ denote the original Mellin variables in the lower-point amplitudes), we obtain a recursive formula for Mellin amplitudes (note that scalar Mellin amplitudes have a similar factorization structure~\cite{Fitzpatrick:2011ia})
\begin{multline}
\label{facMellin}
\mathscr{M}_{n+1}^{\textbf{$\bm{M_i}$}}=\pi^h\int_c\mathscr{M}_3^{M_nM_{n+1}M}(\Delta_n,\Delta_{n+1},h-c)\ \eta_{MN}\\
\overset{\leftrightarrow}{\otimes}\mathscr{M}_n^{M_1M_2\cdots M_{n-1}N}\left(h+c,\Delta_1,\cdots,\Delta_{n-2}\right)\ ,
\end{multline}
where \mbox{$\overset{\leftrightarrow}{\otimes}$} denotes an operator that acts on the Mellin amplitudes and involves an integration over $\gamma_{ij}'$ ($1\le i,j\le n-1$). Since $\Gamma(\gamma_{ij})\to\Gamma(\gamma_{ij}-\gamma_{ij}')$, there are poles at $\gamma_{ij}'=\gamma_{ij}+n_{ij}$ with non-negative integers $n_{ij}$. Integrating around them with the constraints on Mellin variables, one can get a discrete sum of simple pole terms, in the form of
\begin{multline}
\overset{\leftrightarrow}{\otimes}=\sum_{m=0}^\infty\frac{\overset{\leftrightarrow}{\otimes}_m}{\gamma_{n(n+1)}-\frac{\Delta_n^-+\Delta_{n+1}^--h-c}{2}+m}\ ,
\end{multline}
with $\Delta_i^-=\sum_{j\neq i}\gamma_{ij}$ which constrains the Mellin varaiables.\footnote{More specifically, $\Delta_i^-=\Delta_i-\delta_i$ with $\delta_i$ the scaling dimension of $P_i$ in the Mellin amplitude.} More explicitly, the action of $\overset{\leftrightarrow}{\otimes}_m$ is made up of\footnote{Here we suppress indices for notation simplicity.}
\begin{multline}
\mathscr{M}_3\ \overset{\leftarrow}{\otimes}_m=\mathscr{M}_3
\frac{\Gamma\left(\gamma_{n(n+1)}+c+m\right)\Gamma(-c-m)}{\Gamma\left(\gamma_{n(n+1)}\right)}\ ,
\end{multline}
and
\begin{equation}
\overset{\rightarrow}{\otimes}_m\mathscr{M}_n=\sum_{\sum n_{ij}=m}\prod_{i<j}^{n-1}\frac{(\gamma_{ij})_{n_{ij}}}{n_{ij}!}\left.\mathscr{M}_n\right|_{\gamma_{ij}\to\gamma_{ij}+n_{ij}}\ .
\end{equation}
The formal expression (\ref{facMellin}) reduces to strikingly simple expressions for gluon in AdS as we will see in the subsequent sections.
\section{Algorithm for Yang-Mills theory}\label{III}
\begin{figure}
	\centering
	\subfigure[]{
	\begin{minipage}[t]{0.3\linewidth}
	\centering
	\includegraphics[width=1in]{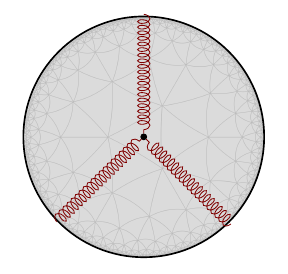}\label{3pt}
	\end{minipage}}
	\subfigure[]{
	\begin{minipage}[t]{0.3\linewidth}
	\centering
	\includegraphics[width=1in]{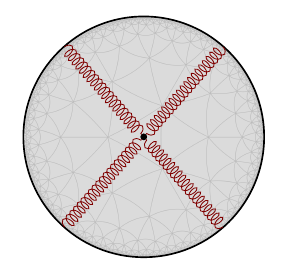}\label{4ptc}
	\end{minipage}}
	\subfigure[]{
	\begin{minipage}[t]{0.3\linewidth}
	\centering
	\includegraphics[width=1in]{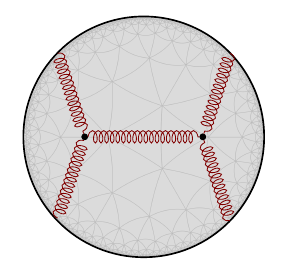}\label{4pts}
	\end{minipage}}
	\centering
	\subfigure[]{
	\begin{minipage}[t]{0.3\linewidth}
	\centering
	\includegraphics[width=1in]{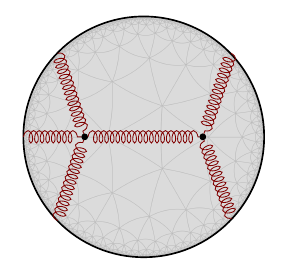}\label{fig:5channel1}
	\end{minipage}}
	\subfigure[]{
	\begin{minipage}[t]{0.3\linewidth}
	\centering
	\includegraphics[width=1in]{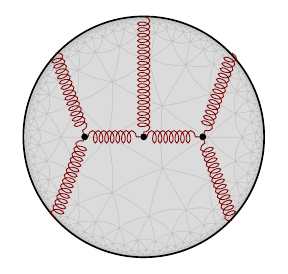}\label{fig:5channel2}
	\end{minipage}}
	\subfigure[]{
	\begin{minipage}[t]{0.3\linewidth}
	\centering
	\includegraphics[width=1in]{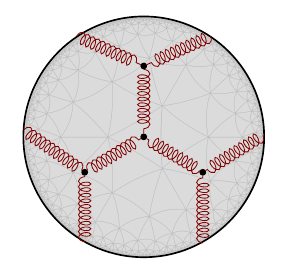}\label{fig:6channel2}
	\end{minipage}}
	\centering
	\subfigure[]{
	\begin{minipage}[t]{0.3\linewidth}
	\centering
	\includegraphics[width=1in]{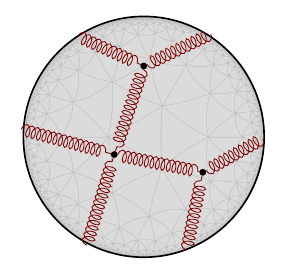}\label{fig:6channel3}
	\end{minipage}}
	\subfigure[]{
	\begin{minipage}[t]{0.3\linewidth}
	\centering
	\includegraphics[width=1in]{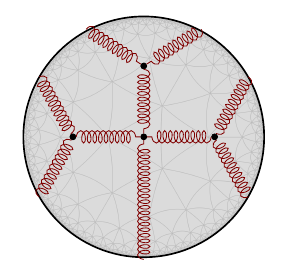}\label{7channel1}
	\end{minipage}}
	\subfigure[]{
	\begin{minipage}[t]{0.3\linewidth}
	\centering
	\includegraphics[width=1in]{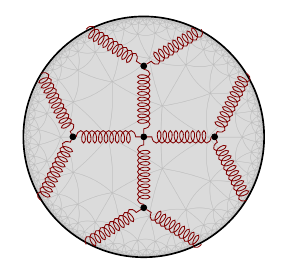}\label{8channel1}
	\end{minipage}}
	\centering
\caption {\label{diagrams} From simple to complex: A compendium of tree-level higher-point Witten diagrams for gluons}
\end{figure}
Here, we explain the main technical details. First, it is convenient to write the Mellin amplitude in terms of a reduced one, namely \mbox{$\pi^h/2 \left(\prod_{i=1}^n C_i \widehat{D}_{\Delta_i}^{M_iA_i}/\ \Gamma(\Delta_i)\right)\tilde{\mathscr{M}}_{n,\textbf{$\bm{A_i}$}}$}.\footnote{Here we suppress the color factor temporarily, since it is a spectator of the subsequent procedure. We will retrieve it in the final result (\ref{nM}) of this section.} We will derive a recursion relation for \emph{the reduced Mellin amplitude} $\tilde{\mathscr{M}}$ in the context of Yang-Mills theory in AdS, starting from (\ref{facMellin}). Recall that this formula is for a diagram factorized into a 3-point amplitude $\mathscr{M}_3$ and an $n$-point amplitude $\mathscr{M}_n$ as in Figure \ref{nchannel}. Here is the algorithm:
\begin{enumerate}
    
\item
\emph{Action of $\widehat{D}$ on $3$-point and simplification of the pole:} For the amplitude of the $3$ points (see Figure \ref{3pt}), calculated in \cite{Paulos:2011ie},
\begin{multline}
\label{eq:3J1}
\tilde{\mathscr{M}}^{a_1a_2a_3}_{3,\textbf{$\bm{A_i}$}}=ig f^{a_1a_2a_3}\Gamma\left(\frac{\sum_{i=1}^3\Delta_i-d+1}{2}\right)\mathscr{I}_{\textbf{$\bm{A_i}$}}
\end{multline}
where $\mathscr{I}_{\textbf{$\bm{A_i}$}} =2\eta_{A_2A_3}(P_2-P_3)_{A_1}+2\eta_{A_3A_1}(P_3-P_1)_{A_2}+ 2\eta_{A_1A_2}(P_1-P_2)_{A_3}$. To be consistent with Fig.\ref{nchannel}, we map $(P_1,P_2,P_3)$ to $(P_n,P_{n+1},Q)$. We want to eliminate the $Q$ dependence in order to use \eqref{facMellin}, which appears in $\tilde{\mathscr{M}}_3$ as well as in $\widehat{D}_{h-c}^{MA}$. Hence, we use the generalized momentum conservation to replace $Q$ by $P_n$ and $P_{n+1}$. As a result, one arrives at a simple expression with the replacement $\widehat{D}_{h-c}^{MA} \mathscr{I}_{A_nA_{n+1}A} \to X_{n(n+1)}^M (h-c-1)/(h-c)$\cite{Kharel:2013mka}, where we have defined
\begin{equation}
\label{Xij}
X_{ij}^{M}\equiv 2\left(\eta_{A_iA_j}P_i^M-2\delta^M_{A_i}P_{i,A_j}\right)-(i\leftrightarrow j)\ .
\end{equation} 
Importantly, the zero $c=h-1$ coincides with one of the double poles in $\int_c$ when $\Delta=d-1$. So, the pole $c=h-1$ becomes simple. And the integration around it can be easily performed.

\item\emph{Action of $\widehat{D}$ on $n$-point:} Similarly, for the $n$-point amplitude, we will label the internal boundary point as $Q$, which we would like to eliminate. We first act $\widehat{D}_{h+c}^{MA}$ on $\tilde{\mathscr{M}}_{n,\textbf{$\bm{A_i}$}A}\prod_{i}(-2P_i\cdot Q)^{-l_i}$. The action of the first term in the differential operator gives an additional factor $\eta^{MA}(h+c-1)/(h+c)$. For the second term in $\widehat{D}^{MA}$, we contract $Q^{A}\tilde{\mathscr{M}}_{\textbf{$\bm{A_i}$}A}$, eliminate the $Q$ dependence again by using the generalized momentum conservation, and perform the action of $\partial_{Q^M}$.\footnote{For $Q$ appearing in product $P_i\cdot Q$, we can simply absorb it into the basis of Mellin space, $(P_i\cdot Q)^{-l_i}$, by shifting $l_i$.} Finally, the action of the second term in $\widehat{D}^{MA}$ gives an additional factor $2\sum_{i=1}^{n-1}P_i^M/(h+c)$.
    
\item\emph{Simplification of $(n+1)$-point:} Now we glue the 3-point to the $n$-point, as in Fig.\ref{nchannel}. Due to the antisymmetry of (\ref{Xij}), we can show that the term with $\sum_{i=1}^{n-1}P_i^M$ does not contribute to the total $(n+1)$-point amplitude (\ref{facMellin}). In (\ref{facMellin}), $\sum_{i=1}^{n-1}P_i^M$ contracts with $X_{n(n+1)}^M$, (\ref{Xij}). The contraction leads to two kinds of terms, with free indices in the metric or in the boundary point. For the former, we can shift the Mellin variables and get $\eta_{A_nA_{n+1}}(\Delta_n-\Delta_{n+1})$, which vanishes when we take $\Delta_i=d-1$. For the latter, we can use the generalized momentum conservation to replace $  P_{n,A_{n+1}} \sum_{i=1}^{n-1}P_{i,A_n}$ by $-P_{n,A_{n+1}}P_{n+1,A_n}$, which is also cancelled out by the antisymmetry between the label $n$ and $n+1$. 
\end{enumerate}

The above algorithm leads to dramatic simplification of (\ref{facMellin}). Finally we have the following refined recursion formula for the reduced Mellin amplitudes
\begin{multline}
\label{nM}
\tilde{\mathscr{M}}_{n+1,{\textbf{$\bm{A_i}$}}}^{\textbf{$\bm{a_i}$}}
= g f^{ba_na_{n+1}} X_{n(n+1)}^M\sum_{m=0}^\infty\frac{\V_3^{m,0,0}}{\tilde{\gamma}_{n(n+1)}}\sum_{\sum_{i<j}^{n-1}n_{ij}=m}\\ m!\prod_{i<j}^{n-1}\frac{(\gamma_{ij})_{n_{ij}}}{n_{ij}!}\left.\tilde{\mathscr{M}}_{n,{\textbf{$\bm{A_i}$}}M}^{{\textbf{$\bm{a_i}$}}b}\right|_{\gamma_{ij}\to\gamma_{ij}+n_{ij}}\ ,
\end{multline}
where we have defined $\tilde{\gamma}_{ij}\equiv 4m!\ \Gamma\left(h+m\right)(\gamma_{ij}-h+m)$ and $\V_3^{m,0,0}\equiv \left(h-m\right)_m\Gamma\left(d-1\right)$ which represents the contribution of the three-vertices connecting to one bulk-to-bulk propagator. Note that with $n=0$, $\V_3^{n,0,0}$ reduces to the expression $\Gamma\left(d-1\right)$, which is the factor that appears in the amplitude of the three gluons (\ref{eq:3J1}) with $\Delta_i=d-1$.

\section{Map to flat amplitudes}\label{IV}
\begin{figure}
	\centering
\includegraphics[scale=0.8]{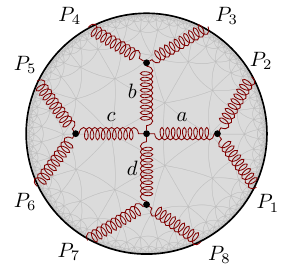}
\caption{\label{drone} Decoding the complexity: topology of eight-point ``Drone diagrams'' as a representative example (\ref{M8channel1})}
\end{figure}

Now equipped with the necessary tools, we are poised to calculate the gluon amplitudes spanning from lower to higher points. We present salient features for various topologies, as illustrated in Figs. \ref{diagrams}\subref{4pts}--\subref{8channel1}. As a non-trivial and representative example, we showcase a result for a higher-point computation in Figure \ref{drone} that hasn't been computed before in literature. We will discuss the detailed computation for this and other higher point topologies in the accompanying paper \cite{accompanying}. 

We will also delineate a simple correspondence between flat space amplitudes and AdS amplitudes for gluons. After rigorous computation of the Mellin amplitudes for the diagrams shown in Figure \ref{diagrams}\footnote{Previous calculations for the 3, 4, and 5-point amplitudes can be found in \cite{Paulos:2011ie,Kharel:2013mka}}, we identify a direct correspondence between these amplitudes (modulo for differential operators $\widehat{D}$) and their flat-space analogs. This correspondence is tabulated in Table \ref{dict}.

\renewcommand{\arraystretch}{1.5}
\begin{table}
\centering
\caption{This table provides a dictionary between flat space and AdS amplitudes}
\label{dict}
\begin{tabular}{lcc}
\hline
\texttt{Description} & \texttt{Flat Space} & \texttt{AdS} \\
\hline
Amplitude denotation & $\mathscr{A}_{n,\textbf{$\bm{A_i}$}}$ & $\tilde{\mathscr{M}}_{n,\textbf{$\bm{A_i}$}}$ \\
Kinematic variable & $ik_i$ & $2P_i$ \\
Internal propagator & $i(2\sum k_i\cdot k_j)^{-1}$ & $(\tilde{\sum}\gamma_{ij})^{-1}$ \\
Three-vertex coupling& $g$ & $g\V_3^{n_a,n_b,n_c}$ \\
Four-vertex coupling & $g^2$ & $g^2\V_4^{n_a,\cdots, n_d}$ \\
\hline
\end{tabular}
\end{table}

As one can see in the dictionary, the kinematic variables, momenta and boundary points, are mapped to each other. In the previous sections we have already seen several examples where these two variables mimic each other. First, they both satisfy the null condition, $P^2=0$ and $k^2=0$. Second, there is a generalized momentum conservation for the boundary points, resembling the conservation law $\sum_ik_i=0$, as discussed in Section \ref{II}. In addition, we can also provide a quantitative understanding as follows. In scattering amplitudes, kinematic variables with free indices arise from the derivative term in the action. For flat space, applying the derivative $\partial_{x^A}$ on $e^{ik_ix}$(where $x$ denotes the interaction vertex location) results in a factor of $ik^A$. For AdS space, the action of $\partial_{X^A}$ on the boundary-to-bulk propagator \mbox{$\mathcal{E}_{\Delta_i}^{M_iA_i}(P_i,X)$} gives a factor of $2P_i^A$. This is why $ik^A$ and $2P_i^A$ appear on the two sides of the dictionary.
However, for the internal propagator, the inner product of momenta is mapped to the Mellin variables. More specifically, each bulk-to-bulk propagator is associated with an integer $m$ to be summed from 0 to $\infty$. And we have defined
\begin{multline}
\tilde{\sum_{i<j}}\gamma_{ij}\equiv 4m!\ \Gamma\left(d/2+m\right)\bigg(\sum_{i<j}\gamma_{ij}+(d-1)/2\\
-\sum_i\left(d-1-\delta_i\right)/2+m\bigg)\ .
\end{multline}
The map between $k_i\cdot k_j$ and $\gamma_{ij}$ is similar to the one discussed in the scalar case \cite{Paulos:2011ie}. This at first may look surprising. However, from the above discussion we can infer that $k_i\cdot k_j$ is mapped to $P_{ij}$. Alternatively, we can shift Mellin variables $\gamma_{ij}\to \gamma_{ij}+1$. Then, the Mellin basis $\prod \Gamma(\gamma_{ij})P_{ij}^{-\gamma_{ij}}$ ``swallows" the $P_{ij}$ and returns a $\gamma_{ij}$.

We also observe an interesting coupling map when studying interaction vertices. This is represented on the AdS side by an additional factor, $\V^{n_a,n_b,\cdots}$, which indicates a vertex connected to propagators indexed by $n_a,n_b,\cdots$ integers. Notably, this includes boundary-to-bulk propagators with $n=0$. The definitions of $\V_3$ and $\V_4$ are provided in Appendices \ref{V3} and \ref{V4}.

Utilizing the supplied map, we present an intricate illustration of the eight-point Mellin amplitude within the Drone channel, as showcased in Figure \ref{drone}
\begin{multline}
\label{M8channel1}
\mathscr{M}_{\texttt{Drone}}^{M_1M_2\cdots M_8}
=g^6 \frac{\pi^h}{2} \left(\Di{8} \right)  \\ \times\sum_{{\bf n}=0}^\infty
\frac{\V_3^{n_a,0,0}}{\tilde{\gamma}_{12}(n_a)}\frac{\V_3^{n_b,0,0}}{\tilde{\gamma}_{34}(n_b)} \V_4^{n_a,n_b,n_c,n_d}  \frac{\V_3^{n_c,0,0}}{\tilde{\gamma}_{56}(n_c)}\frac{\V_3^{n_d,0,0}}{\tilde{\gamma}_{78}(n_d)}\\
\times \left(X_{12}^{N_a}~X_{34}^{N_b}~X_{56}^{N_c}~X_{78}^{N_d}\right)f^{aa_1a_2}f^{ba_3a_4}f^{ca_5a_6}f^{da_7a_8}\\
\times\big((f^{acb'}f^{bdb'}+f^{adb'}f^{bcb'})\eta_{N_aN_b}\eta_{N_cN_d}\\
+\text{cyclic perm. of }(b,c,d)\big)\ ,
\end{multline}
where, ${\bf{n}}=\{ n_a,n_b,n_c,n_d\}$ and $X_{ij}$ is defined in \eqref{Xij}. One can see how closely it resembles its flat-space counterpart. We see similar patterns for the other topologies in Figure \ref{diagrams} by explicit computation~\cite{accompanying}. We are tempted to conjecture, based on the results of our complicated computation, that it works for any tree-level gluon amplitude in AdS. 

Finally, it is worth mentioning that in the high energy limit $\gamma_{ij}\to \infty$, the scalar Mellin amplitude reduces to the flat-space amplitude \cite{Penedones:2010ue}. This flat-space limit is further proven in~\cite{Fitzpatrick:2011ia}. We now generalize it to spinning cases
\begin{multline}
\label{Mnflat}
\tilde{\mathscr{M}}_{{\textbf{$\bm{A_i}$}}}\approx \int_0^\infty d\beta\beta^{\frac{\sum_{i=1}^n\Delta_i-d}{2}-1}e^{-\beta} \mathscr{A}_{{\textbf{$\bm{A_i}$}}}\bigg(ik_i\to 2\sqrt{\beta} P_i,\\
 i(2\sum k_i\cdot k_j)^{-1}\to (4\beta\sum \gamma_{ij})^{-1}\bigg)\ .
\end{multline}
For more details see~\cite{accompanying}.

\section{Discussion and Future Directions}\label{V}
In this work, we have explored a technique to recursively compute tree-level gluon amplitudes in AdS in Mellin space. Our analysis of several higher-point amplitudes reveals a noteworthy resemblance to flat-space counterparts, paving the way for us to write a dictionary between tree-level amplitudes in AdS and amplitudes in flat space for gluons. 

This dictionary opens several avenues for future research. Of particular interest is the application of our techniques to higher-spin particles, such as gravitons, which show promise for computation with analogous methods. Further, extending and using the dictionary to compute loop-level AdS computations presents a significant opportunity (see some advances in spinning loop \cite{Giombi:2017hpr,Albayrak:2020bso,Alday:2021ajh}). 
Additionally, the flexibility of our methods indicates potential for broader applications, extending to the cosmological bootstrap program and computations of de Sitter space correlators — fields that stand to benefit greatly from the development of spinning Mellin space technology \cite{Arkani-Hamed:2018kmz,Baumann:2019oyu,Baumann:2020dch,Sleight:2019hfp, Sleight:2021plv}.

With the proposed dictionary, we can tackle several important research problems in this area. It would be interesting to develop a refined version of the flat-space BCFW relations to calculate higher-point AdS amplitudes. The efficacy of these methods in momentum space for gluons and graviton amplitudes is well established\cite{Raju:2012zs,Raju:2012zr,Albayrak:2023jzl}; however, extending this success to calculations beyond four-point remains a formidable challenge. The succinctness of the results and a clear map to flat space suggest a possibility of a generalized BCFW in Mellin space. 

Another significant avenue for further research is the explicit construction of the double copy framework in Anti-de Sitter (AdS) spaces. Although our grasp of perturbative gravitational dynamics in curved spacetime is still rudimentary compared to that in flat space, there have been promising developments\cite{Albayrak:2020fyp, Armstrong:2020woi, Farrow:2018yni, Diwakar:2021juk, Zhou:2021gnu, Jain:2021qcl, Alday:2022lkk, Herderschee:2022ntr, Cheung:2022pdk}. These methodologies, while innovative, are limited to lower point functions. One exception is the color/kinematics duality construction in five-point functions for supersymmetric theories within AdS\(_5\) \cite{Alday:2022lkk}. With the application of our dictionary, we anticipate that the complexities associated with formulating double copy can be considerably reduced, aligning AdS amplitude computations with their flat-space counterparts. We are actively pursuing this line of research.
\begin{acknowledgments}
We want to thank Soner Albayrak and Xinkang Wang for discussions. 
\end{acknowledgments}
\appendix
\renewcommand{\theequation}{\thesection.\arabic{equation}}

\section{Definition of $\V_3$}\label{V3}
We start from the definition of $ \V_3^{m,n,0}$.
\begin{multline}
\label{V3mn0}
\V_3^{m,n,0}\equiv \Gamma(d-1)\big(\frac{d}{2}-m+n\big)_m\big(\frac{d}{2}-n+m\big)_n\ .
\end{multline}
This definition is explicitly symmetric under $m\leftrightarrow n$, and when $m=0$ it is consistent with $\V_3^{n,0,0}$ defined before. 

Then, we can present the definition of $\V_3^{m_1,m_2,m_3}$,
\begin{multline}
\label{V33}
\V_3^{m_1,m_2,m_3}\equiv \sum_{n_2=0}^{\min\{m_3,m_2\}}\sum_{n_1=0}^{\min\{m_3-n_2,m_1\}}\frac{m_3!}{n_1!n_2!}\\
\times \frac{\left(\frac{d}{2}-m_3+m_1+m_2\right)_{m_3-n_1-n_2}}{(m_3-n_1-n_2)!}\V_3^{m_1-n_1,m_2-n_2,0}\\
\times\prod_{i=1}^2(m_i-n_i+1)_{n_i}\left(\frac{d}{2}+m_i-n_i\right)_{n_i}\ .
\end{multline}

Presumably this definition is symmetric among $m_i$, for which we do not have a proof. But we can check from the definition (\ref{V33}) that $\V_3^{m,n,0}=\V_3^{m,0,n}=\V_3^{0,m,n}$.
\section{Definition of $\V_4$}\label{V4}
We start from
\begin{multline}
\label{V4mn00}
\V_4^{m,n,0,0}\equiv \Gamma\left(\frac{3d-4}{2}\right)m!\sum_{n_1=0}^{\min\{m,n\}}\\
\frac{(d-1-m+n)_{m-n_1}}{n_1!(m-n_1)!}(n-n_1+1)_{n_1}\\
\times\left(\frac{d}{2}+n-n_1\right)_{n_1}(d-1-n+n_1)_{n-n_1}\ .
\end{multline}
Then,
\begin{equation}
\label{V4mnm0}
\begin{split}
&\V_4^{m_1,m_2,m_3,0}\equiv m_3!\sum_{n_2=0}^{\min\{m_3,m_2\}}\sum_{n_`=0}^{\min\{m_3-n_2,m_1\}}\\
&\times\frac{(d-1-m_3+m_1+m_2)_{m_3-n_1-n_2}}{n_1!n_2!(m_3-n_1-n_2)!}\V_4^{m_1-n_1,m_2-n_2,0,0}\\
&\times\prod_{i=1}^2(m_i-n_i+1)_{n_i} \left(\frac{d}{2}+m_i-n_i\right)_{n_i}\ .
\end{split}
\end{equation}
One can check that from this definition, $\V_4^{m,n,0,0}=\V_4^{0,m,n,0}=\V_4^{m,0,n,0}$ and reduces to $\V_4^{m,n,0,0}$ in (\ref{V4mn00}).

Finally,
\begin{equation}
\label{V4mnnm}
\begin{split}
&\V_4^{m_1,m_2,m_3,m_4}\equiv m_4!\sum_{n_2=0}^{\min\{m_4,m_2\}}\sum_{n_1=0}^{\min\{m_4-n_2,m_1\}}\\
&\sum_{n_3=0}^{\min\{m_4-n_1-n_2,m_3\}}\V_4^{m'-n_{34},n-n_{12},n'-n_{56},0}\\
&\times\frac{(d-1-m_4+m_1+m_2+m_3)_{m_4-n_1-n_2-n_3}}{n_1!n_2!n_3!(m_4-n_1-n_2-n_3)!}\\
&\times \prod_{i=1}^3(m_i-n_i+1)_{n_i}\left(\frac{d}{2}+m_i-n_i\right)_{n_i}\ .
\end{split}
\end{equation}
One can check that from this definition $\V_4^{m_1,m_2,m_3,0}=\V_4^{m_1,m_2,0,m_3}=\V_4^{m_1,0,m_2,m_3}=\V_4^{0,m_1,m_2,m_3}$ and reduces to (\ref{V4mnm0}).

\bibliography{savanreference}
\end{document}